\documentclass{iau}

\usepackage{amsmath}
\usepackage{graphicx}
\usepackage{multirow}

\newcommand {\Msun} {M$_\odot$}
\newcommand {\Reff} {R$_\mathrm{e}$}
\newcommand {\sersic} {S\'{e}rsic }

\begin{document}

\lefttitle{Ver\v{s}i\v{c} et al.}
\righttitle{IAU Symposium 379: Flattening of dark matter halo of NGC~5128}

\jnlPage{1}{7}
\jnlDoiYr{2023}
\doival{10.1017/xxxxx}

\aopheadtitle{Proceedings of IAU Symposium 379}
\editors{P. Bonifacio,  M.-R. Cioni, F. Hammer, M.S.  Pawlowski \& S. Taibi,  eds.}

\title{The flattening of dark matter halo of Centaurus A galaxy (NGC~5128) out to 40 kpc}

\author{T. Ver\v{s}i\v{c}$^{1,2}$
, M. Rejkuba$^2$, M. Arnaboldi$^2$, O. Gerhard$^3$, J. Hartke$^{4,5}$, C. Pulsoni$^3$, G. van de Ven$^{1}$}
\affiliation{
$^1$ Department of Astrophysics, University of Vienna, Türkenschanzstrasse 17, A-1180 Vienna, Austria
email:~\email{tadeja.versic@univie.ac.at}\\
$^2$ European Southern Observatory, Karl-Schwarzschild-Str. 2, 85748 Garching, Germany\\
$^3$ Max-Planck-Institut für extraterrestrische Physik, Giessenbachstraße, 85748 Garching, Germany\\
$^4$ Finnish Centre for Astronomy with ESO (FINCA), University of Turku, FI-20014 Turku, Finland \\
$^5$ Tuorla Observatory, Department of Physics and Astronomy, FI-20014 University of Turku, Finland}

\begin{abstract}
Cosmological simulations predict dark matter shapes that deviate from spherical symmetry. The exact shape depends on the prescription of the simulation and the interplay between dark matter and baryons. This signature is most pronounced in the diffuse galactic haloes that can be observationally probed with planetary nebulae and globular clusters (GCs). The kinematic observations of these halo tracers support intrinsic triaxial shape for the mass generating the gravitational potential. 
With discrete axisymmetric modelling of GCs as the halo tracers of NGC~5128 we investigate the overall mass distribution of this nearby giant elliptical galaxy. Our modelling approach constrains $c_{200}$, $(M/L)_{\star, B}$ and inclination.
We derive a preliminary $M_{200}\sim 1 \times 10^{12}$ \Msun\  and flattening $q_{\mathrm{DM}}\sim 1.3$ indicative of prolate/triaxial halo for NGC~5128.
\end{abstract}


\maketitle

\section{Introduction}

Mass is the dominant driver of structure formation and evolution.
In the concordance model of $\Lambda$CDM cosmology, dark matter is the dominant mass component in the Universe.
$\Lambda$CDM has been very successful at explaining the observed properties of galaxy populations. 
However, there are still some predictions from the theory that have been challenging to constrain observationally.
One such challenge is the intrinsic shape of the dark matter haloes.
The shape is quantified with the axis ratios of a 3D ellipsoid: $p = b/a$ and $q = c/a$, where $a>b>c$ are semi-major, intermediate and minor axis.
Halos with $a>b=c$ are considered prolate and $a=b>c$ are considered oblate, otherwise, the shape is triaxial.

NGC~5128 (CenA) is a nearby early-type galaxy (ETG), owing to its proximity and mass it is an excellent target for dynamical modelling of its halo tracers to constrain the flattening of the dark matter halo.
At a distance of 3.8 Mpc \citep{Harris+10_distance} and with a stellar mass of $M_\star\sim10^{11}$\Msun\ this galaxy hosts
1700 discrete kinematic tracers, Globular Clusters (GCs) and Planetary Nebulae (PNe) with velocity measurements in the literature.
Using PNe kinematics \cite{Peng+04_PNe_dyna_outer_halo} found evidence for a triaxial potential of CenA.

In this proceeding, we outline the scientific rationale behind the dynamical studies of dark matter shapes, introduce  NGC~5128  as a test case and discuss the first steps of the dynamical modelling of its halo tracers.

\section{Dark matter halo shapes}

Observationally, dark matter can only be probed through its gravitational influence on the baryons.
We can use dynamical modelling to investigate the distribution of dark matter that best explains the motions of luminous matter in the dark matter-dominated galactic haloes.
To carry out dynamical modelling and measure the shape of dark matter at these radii we need kinematic tracers of the galactic halo: GCs and PNe.
In cosmological simulations, on the other hand, the distribution and shape of dark and baryonic matter can be determined directly through the positions of the simulated particles.

\subsection{Simulations}

Studies of N-body dissipationless simulations \citep{Allgood+06} found a strong dependence of the halo shape with the different cosmological parameters and the dark matter fraction. 
These authors find dark matter haloes are prolate in the centres and become more spherical close to the virial radius, and similar conclusion was reached by \cite{Hayashi+07}.
In the Illustris simulations, \cite{Chua2019_halo_shapes} found that baryonic physics acts to make the haloes rounder and more oblate.
The impact varies with the halo mass and has strong radial variation, the strongest impact is for $M_{200} = 10^{12-13}$M$_\odot$ at $r\sim 0.15r_{200}$.

\subsection{Observational constraints}

The Milky Way provides a unique vantage point and the possibility to study a galaxy in the greatest detail.
However, depending on the technique and data used the recovered shape of the dark matter halo of the Galaxy vary \citep[][and references therein]{BlandHawthorn+16_MW_review}.

Beyond the Milky Way isolated studies have found strongly prolate haloes for WLM, a gas-rich dwarf galaxy with the major axis of the dark and baryonic components perpendicular to each other \citep{Leung+19}.
\cite{Khoperskov+14} found that polar ring galaxies are hosted in haloes that are more prolate in the outer regions and oblate in the centres.
While dynamical modelling enables direct measurements of the shape, \cite{Pulsoni+18} find that ETGs have more diverse kinematic properties in their halos than in the central regions. With PNe they find all slow rotators and 40\% of fast rotators show signatures of triaxial halos such as kinematic twists or misalignments.  

\section{The Centaurus A galaxy}

NGC~5128 is the closest early-type galaxy to us classified as giant Elliptical or S0 and has clear evidence of a recent accretion history or possibly a major merger \citep[][and references therein]{Wang+20_CenAmergerhistory}.
It hosts a radio source Cenutarus A and for simplicity, we will refer to the galaxy as CenA.
Its proximity has enabled detailed studies of the resolved RGB (red giant branch) stars out to more than 25 \Reff (140 kpc) in \cite{Rejkuba+22} providing constraints on the stellar halo distribution.
Spectroscopic observations of a large population of PNe \citep{Peng+04_PNe_dyna_outer_halo, Walsh+15} and GCs \citep{Hughes+23} over a vast halo area provide a necessary sample of kinematic tracers for the dynamical modelling that we are exploiting in this work.

\subsection{Kinematics of the globular clusters}

\begin{figure}
  \centering
\includegraphics[scale=.3]{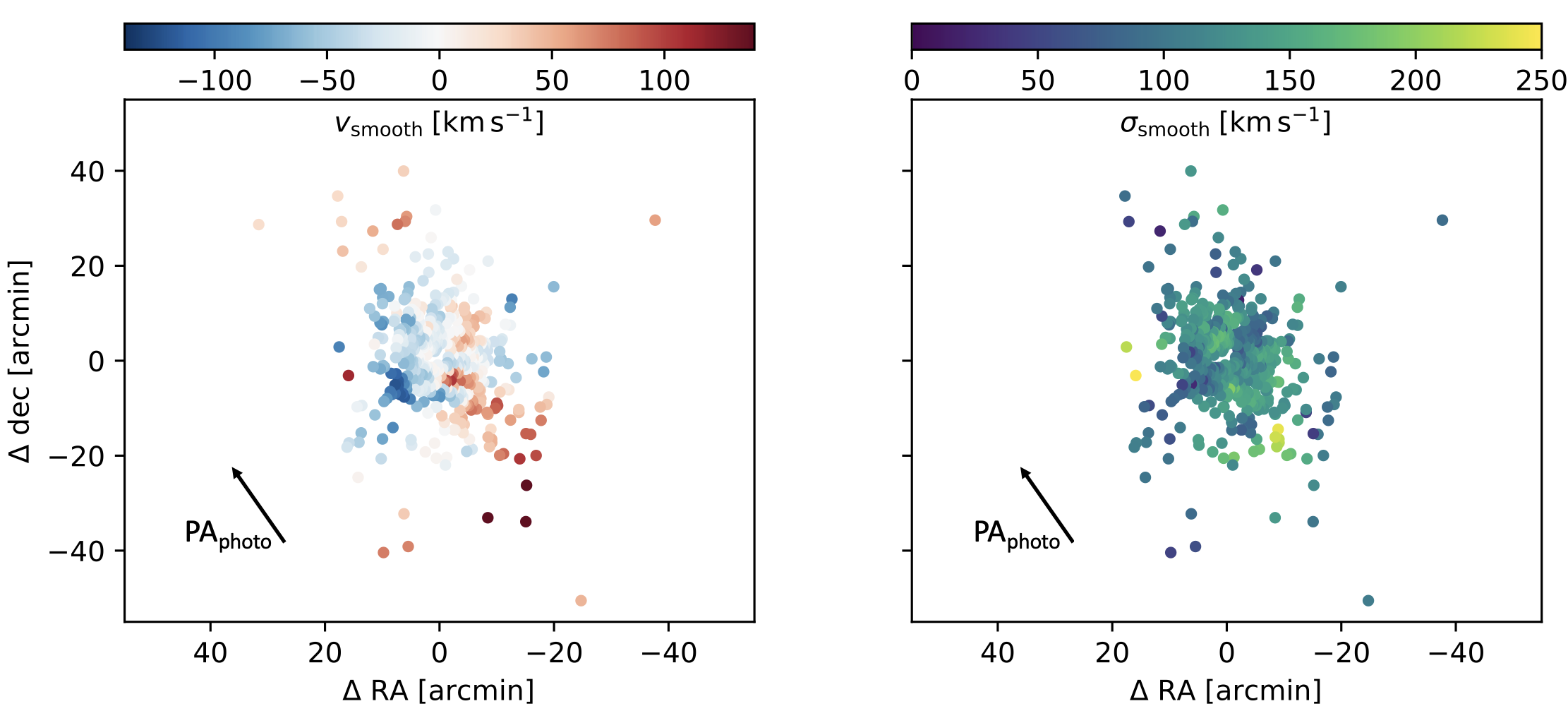}
  \caption{Smooth velocity  and velocity dispersion field of GCs in CenA on the left and the right panel respectively. The photometric position angle of the stellar component is indicated with the arrow.}
  \label{fig:smooth_velocity_field}
\end{figure}

In this work, we use the kinematic sample compiled from the literature by \cite{Hughes+21} of around 550 GCs out to a median distance of 40 kpc.
To investigate the velocity field of the GCs used in this work we adopted the Gaussian smoothing approach from \cite{Pulsoni+18} tested on simulated galaxies for a similarly sized discrete sample of kinematic tracers.
Figure~\ref{fig:smooth_velocity_field} shows the smooth velocity on the left panel and velocity dispersion on the right of the GCs within 50 arcmin ($\sim 10$ \Reff).
We can see signs of rotation both along the major and minor axis similar to the PNe velocity field.
The velocity dispersion field shows a drop in the velocity dispersion along the minor compared with the major axis.

The GCs of CenA show bimodal colour distribution with red GCs exhibiting similar velocity properties as the field stars, traced by the PNe and blue GCs showing weaker rotation \citep[e.g.][]{Peng04_GC_age_Z_kin}.
\cite{Woodley+10_ages_Z_GC} confirmed the presence of metallicity bimodality first identified as a colour bimodality.
Additionally, they found that $\sim50\%$ of metal-rich GCs formed in the past 8Gyrs, supporting the continuous accretion and mergers in CenA.

\subsection{Kinematics of the planetary nebulae}

\cite{Peng+04_PNe_dyna_outer_halo} found that the inner rotation of the PNe along the major axis of the galaxy is extended out to the limit of their data (80kpc).
They found evidence of a twist in the zero velocity curve which points to a triaxial or prolate potential of the galaxy.
Using harmonic expansion \cite{Pulsoni+18} provided a systematic study of the PNe kinematics and computed the properties of a smooth velocity field from the discrete sample.
The authors determined the radial variation of the kinematic position angle of the PNe and a drop in the velocity dispersion around 3 \Reff ($\sim$ 15 kpc).

\subsection{Comparison between GC and PNe velocity field}

To investigate the possible presence of substructures we use the smooth velocity field of the PNe from \cite{Pulsoni+18}.
We find the discrete sample of GCs is consistent with being drawn from the smooth velocity field of the PNe.
The 1, 2, and 3$\sigma$ outliers show no preferred position in the plane of the galaxy that would indicate a presence of a dynamically cold component that has recently been accreted.
Such a component would not be dynamically relaxed and would bias dynamical modelling that is based on a steady-state assumption.

\section{Dynamical modelling}

We use \textsc{CJAM} \citep{Watkins+13} implementation of Jeans Anisotropic MGE (Multi Gaussian Expansion), axisymmetric Jeans equations with the assumptions on the cylindrical alignment of the velocity anisotropy introduced by \cite{Cappellari+08}.

\subsection{Introduction to the discrete dynamical modelling}

We assume that a local line-of-sight velocity distribution (LOSVD) can be well described by a Gaussian function and use it as likelihood in our Bayesian analysis.
Therefore, each kinematic tracer is used discretely to inform the likelihood to find the the best-fit parameters. 
This methodology has been used in the literature to investigate enclosed mass profiles, orbital properties of the kinematic tracers and flattening of the dark matter haloes on real and simulated galaxies 
\citep[e.g. ][]{Zhu+16, Hughes+21_dyn_modeling, Leung+19}.

\subsection{Globular cluster tracer density profile}

The largest and most homogenous photometric catalogue of GC candidates is presented in \cite{Taylor+17} and we used it in this work to build the density profile of the kinematic tracers.
To identify a sample of relaxed high-fidelity GCs from the GC candidates we analyzed the spatial symmetry of different magnitude selected populations. 
A dynamically relaxed population is expected to show a point-symmetric spatial distribution. To ensure well-characterized completeness we also require its magnitude distribution to be consistent with a theoretical Gaussian GC luminosity function.

We identified a population, which we called \emph{intermediate population}: $-8.2 < R_0 \leq -7.6$ in the reddening corrected R-band magnitude.
This population shows strong colour bimodality and we determine the colour separation of $(u-z)_0\geq$ 2.7 for the red population and $(u-z)_0<$ 2.7 for the blue, with approximately the same number of GCs in both.
We computed the flattening of both red and blue GCs separately and found blue GCs are rounder.
Finally, we binned them in elliptical radii and found the spatial distribution of GCs can be best described with a combination of an inner \sersic and outer power-law profile.

\begin{figure}[h]
  \centering
\includegraphics[scale=.35]{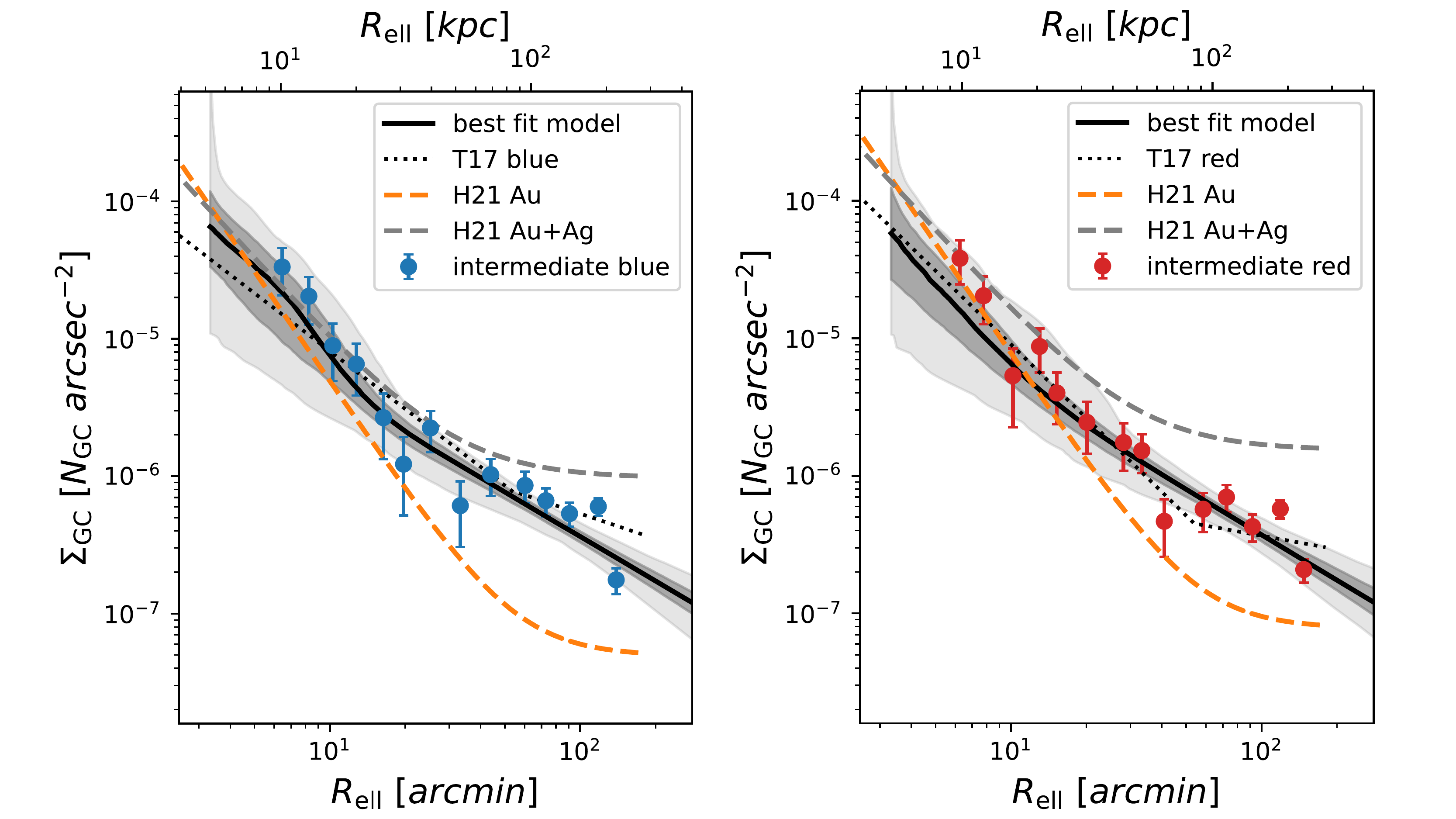}
  \caption{The tracer density profile for the red and blue globular clusters. The left panel shows the best-fit tracer density profile for the blue globular clusters and the right for the res globular clusters.  Black solid lines show the best-fit models with dark and light grey showing 1 and 3 $\sigma$ uncertainties. Dashed and dotted lines show literature results from \cite{Taylor+17}  and \cite{Hughes+21}.}
  \label{fig:tracer_density}
\end{figure}

Figure~\ref{fig:tracer_density} shows the best-fit results for the blue GCs on the left and the red GCs on the right panel.
The data is shown with colored points, the black line shows the best-fit model with dark and light grey 1 and 3$\sigma$ uncertainties.
To compare with the literature, the dotted black line shows the best-fit profiles for red and blue GCs from \cite{Taylor+17} work and the dashed orange and grey line show the golden and silver sample from \cite{Hughes+21}, where different datasets were used to analyze the contamination of the \cite{Taylor+17} GC candidates.
Our profile is broadly consistent with the literature.
In the inner regions, our profiles are steeper than the \cite{Hughes+21} silver and \cite{Taylor+17}  and slightly shallower than the \cite{Hughes+21} golden sample.
In the outer regions, our tracer density drops faster than in all literature studies.

\subsection{Gravitational potential}

In our dynamical modelling, both the stars and dark matter contribute to the gravitational potential.

Using photographic plates, \cite{Dufour79} found the de Vaucouleurs profile can best reproduce the stellar surface density on nested ellipses with a very mild flattening of 0.93 within 1\Reff.
\cite{Rejkuba+22} extended the stellar surface density profile out to 140kpc using resolved observations of RGB stars from HST observations. 
Halo stars were found to have a much flatter distribution with a flattening of 0.54. 
We used both results to create a 2-dimensional surface brightness profile in the range of 0.3$<$R/\Reff$<$25 accounting for the difference in the flattening of central and halo stellar components.

For the dark matter component, we assumed an NFW profile parametrised with the $M_{200}$ and $c_{200}$ parameters.
For each set of $M_{200}$ and $c_{200}$ we decompose the profile with MGE and account for the assumed flattening of the dark matter and inclination.
We define the flattening of the dark matter $q_{\mathrm{DM}} =c/a$, where a is aligned along the photometric major axis and c along the photometric minor axis.

\subsection{Likelihood analysis}

Our model has 7 free parameters: $M_{200}$, $c_{200}$, $q_{\mathrm{DM}}$, inclination, $(M/L)_{\star, \mathrm{B}}$, velocity anisotropy and rotation ($\beta_{z, \mathrm{GC}}$ and $\kappa_{\mathrm{GC}}$ as defined in \cite{Cappellari+08}).
For each combination of parameters, we use \textsc{CJAM} to compute the first and second velocity moments at the position of each GC and a Gaussian LOSVD as the likelihood to determine the best-fit parameters in a discrete approach.
We model separately the red and blue GC populations.

Our modelling approach constrains $c_{200}$, $(M/L)_{\star, B}$ and inclination.
We derive a preliminary dark matter mass $M_{200}\sim 1 \times 10^{12}$ \Msun\ 
and flattening $q_{\mathrm{DM}}\sim 1.3$.
The resulting maximum likelihood solutions for the variables of the gravitational potential are consistent between the blue and red GCs.
With $\kappa_{\mathrm{GC}}$ and $\beta_{z, \mathrm{GC}}$ showing expected differences for both populations of GCs that reflect their distinct dynamic origin.

\section{Conclusions}

The advancement in our understanding of galaxy formation and evolution as well as the increase in the size and quality of the photometric and spectroscopic datasets has made it necessary to move away from spherical symmetry assumptions in dynamical modelling.
With large spectroscopic datasets of halo tracers we can constrain the triaxiality of dark matter.

We find that the magnitude selected intermediate population from \cite{Taylor+17} sample of GCs shows point symmetric spatial distribution, evidence that this population is in approximate dynamical equilibrium.
We find that both red and blue GCs have flattened distributions.
We used this information to construct a tracer density profile that is broadly consistent with the literature but has a declining density profile in the outskirts of the galaxy.
Discrete axisymmetric dynamical modelling 
can constrain the flattening of the dark matter potential out to 40 kpc in CenA.
Detailed results will be presented in Ver\v{s}i\v{c} et al. 2023 in prep.

\section*{Questions}

\textbf{Jianling WANG}:      This is nice work.  I am concerned about the systematics.  The underlying assumption about the dynamics modelling is that the system
is 
at equilibrium state and well relaxed, and single dynamical population
tracers used.  But it is well known that CenA has experienced a
recent major merger, which could violate these assumptions, then what
could be the systematic errors? And it would be useful to apply your
modelling to a realistic model, for example (major merger models for
CenA) to quantify the systematics.

\textbf{Answer}: 
Given that CenA shows evidence of recent accretion the presence of a non-relaxed component is of course a concern. We investigated the potential presence of a non-relaxed and kinematically cold component of our GCs sample with the smooth velocity field determined from the PNe. We found no evidence of a kinematically cold component in the spatially limited sample of GCs we used in this work.  
We, therefore, did not make further analysis of what the systematic errors might be in the presence of strong non-relaxed components.
Literature studies have looked at how the enclosed mass measurements are biased in the presence of non-relaxed tracers \citep[e.g.][]{Alabi+16}.
But I am not aware of a study looking at the impact on the flattening of the potential.
It would be very interesting to investigate and quantify this effect with the simulations as you mentioned.
Thank you for the suggestion.

\section*{Acknowledgements}

\noindent TV \& GvdV acknowledge funding by ERC under EU H2020 grant No. 724857 (ArcheoDyn).

\end{document}